\documentclass[journal]{IEEEtran}

\usepackage{enumerate}
\usepackage{graphicx}
\usepackage{epsf}
\usepackage{amsmath}
\usepackage{amssymb}
\usepackage{array}
\usepackage{setspace}
\usepackage[amsmath,thmmarks]{ntheorem}
\usepackage[ruled,lined,linesnumbered]{algorithm2e}
\usepackage{color}
\usepackage{amsfonts}
\usepackage{dsfont}
\usepackage{xfrac}
\usepackage{breqn}
\usepackage{bbm}
\usepackage{cite}
\usepackage{epstopdf}
\usepackage{amsmath,amssymb}
\usepackage[caption=false,font=footnotesize]{subfig}

\graphicspath{{Fig/},{fig/}}

\theoremseparator{.}

\begin{document}

\title{\Large FSIM: Fluid Element Stacked Intelligent Metasurface for Multiuser Downlink Networks}


\author{
Tsung-Yu Wei, Li-Hsiang Shen, Kai-Ten Feng, and Lie-Liang Yang
}


\maketitle

\begin{abstract}
A fluid element (FE)-aided stacked intelligent metasurface (FSIM) for multiple-input single-output (MISO) communication system is investigated, where a multi-antenna base station (BS) serves multiple single-antenna users through FSIM. Unlike conventional SIM with fixed meta-atom deployment, the architecture allows the meta-atoms in each layer to move within a predefined fluidic region to further increase the spatial diversity. By jointly optimizing the two-dimensional meta-atom positions, BS transmit beamforming, and FSIM phase-shifts, the cascaded BS-FSIM-user channels can be flexibly reconfigured to enhance the desired signals and suppress multiuser interference. The proposed sum-rate maximization problem is highly non-convex and nonlinear due to the coupled solutions of element position, beamforming, and phase-shift. To address this challenge, an alternating optimization (AO) algorithm is developed to iteratively update these variables. The beamforming and FSIM phase-shift subproblems are transformed into semi-definite programming problems and solved by using successive convex approximation (SCA), first-order Taylor approximation, and penalty-based rank-one relaxation, whilst the FE position subproblem is handled through a projected gradient-based update. Simulation results reveal that a compact FSIM with a small inter-layer thickness is preferable, as increasing the thickness weakens inter-layer coupling and degrades the achievable sum rate. Results also demonstrate that the proposed FSIM significantly outperforms conventional SIMs with fixed positions, patch-based structures, partial fluidity, restricted fluid regions, and existing flexible intelligent metasurfaces. Furthermore, the proposed AO-based algorithm achieves superior rate performance compared to sub-schemes, metaheuristic methods, and conventional beamforming benchmarks.

\end{abstract}

\begin{IEEEkeywords}
Stacked intelligent metasurface, fluid element, RIS, alternating optimization, successive convex approximation.
\end{IEEEkeywords}

{\let\thefootnote\relax\footnotetext
{Tsung-Yu Wei and Li-Hsiang Shen are with the Department of Communication Engineering, National Central University, Taoyuan 320317, Taiwan. (email: tywei@g.ncu.edu.tw, shen@ncu.edu.tw)

Kai-Ten Feng is with the Department of Electronics and Electrical Engineering, National Yang Ming Chiao Tung University, Hsinchu 300093, Taiwan. (email: ktfeng@nycu.edu.tw)

Lie-Liang Yang is with the Department of Electronics and Computer Science, University of Southampton, Southampton SO17 1BJ, U.K. (e-mail: lly@ecs.soton.ac.uk)
}
}

\section{Introduction}

Recently, flexible metasurfaces have been proposed as a promising technology to realize smart radio environments \cite{1, my2}. By adjusting the phase-shifts of passive reflecting elements, metasurfaces can reconfigure wireless propagation channels and improve the received signal quality. However, conventional metasurfaces, such as reconfigurable intelligent surfaces (RISs), are generally implemented as a single layer, where the incident signal is controlled by only one set of passive phase-shift coefficients. Such a single-stage reconfiguration structure provides limited flexibility for constructing more flexible channel responses, especially in the multiuser systems where both signal enhancement and interference suppression are required. To overcome this limitation, stacked intelligent metasurface (SIM) has been introduced as a multi-layer metasurface architecture. In paper \cite{3}, SIM as a transmissive RIS structure is proposed for efficient holographic communications, where multiple metasurface layers are utilized to enhance signal processing capability. Furthermore, SIM-aided downlink beamforming has demonstrated that cascaded metasurface layers can provide additional degrees of freedom for enhancing multiuser performance \cite{4}. Although SIM can provide additional spatial diversity through cascaded layers, most existing assumption is the fixed meta-atoms.

Inspired by fluid antenna (FA) system \cite{5}, the fluid element (FE) system is proposed to exploit the elementwise spatial diversity within layered arrays by allowing the position to be adjusted within a given region \cite{6}. 
In \cite{7}, FE multiple access is investigated, showing that position flexibility can create favorable channel conditions and reduce the effect of interference.
Furthermore, RIS-based channel reconfiguration and FA position flexibility are jointly considered to improve communication reliability \cite{8}. Movable-element RIS-aided wireless communication is investigated in \cite{9, my1, 10}, where the element positions are optimized to create favorable channel conditions to improve the rate. In \cite{11}, a RIS-aided FA array-mounted non-terrestrial network is studied, where the FA equipped on the aerial platform provides dynamic position adjustment in addition to RIS channel reconfiguration. A liquid intelligent metasurface-assisted multi-user system is proposed in \cite{12}, where the base station (BS) is equipped with fluid antennas and the metasurface is equipped with liquid elements, which are jointly optimized with beamforming and phase-shift control to improve the sum-rate. RIS-aided wireless communication with movable elements is studied in \cite{13}, where the movement of metasurface elements changes the effective geometry and improves the outage performance compared to conventional RIS. A movable antenna-enabled RIS-aided integrated sensing and communication system is investigated in \cite{14}, where the antenna positions, RIS reflecting coefficients, and BS transmit covariance matrix are jointly optimized to enhance communication and sensing performance. Fluid reconfigurable intelligent surfaces are proposed in \cite{15}, where each fluid element is realized by a dense matrix of subelements, and the selected subelements are jointly designed with discrete phase-shifts to maximize the achievable rate.

Motivated by existing works, incorporating positional reconfigurability into SIM meta-atoms is capable of offering additional spatial diversity for favorable channel conditions than conventional SIM and RISs. This paper proposes an FE-aided SIM (FSIM) communication system, where a multi-antenna BS serves multiple single-antenna users through an FSIM. The meta-atoms in each FSIM layer are allowed to move fluidically within a predefined two-dimensional region for obtaining favorable channel conditions. The BS transmit beamforming, the phase-shifts as well as meta-atom positions of the FSIM are jointly optimized to maximize the system sum-rate. To solve the non-convex problem, an alternating optimization (AO) algorithm is developed. Specifically, the beamforming and phase-shift subproblems are transformed into successive convex approximation (SCA)-based semidefinite programming problems with rank-one relaxation. The position subproblem is handled by a gradient-based update under the movement region and minimum distance constraints. Simulation results demonstrate that the proposed FSIM outperforms the conventional SIMs with fixed positions, patch-based structures, partial fluidity, and restricted fluid regions as well as the existing flexible intelligent metasurface designs in the open literature. Moreover, the proposed AO-based algorithm achieves the highest rate among the considered sub-schemes, metaheuristic methods, and conventional beamforming benchmarks.

\section{System Model and Problem Formulation}
\label{sec_sys}

	We consider a downlink BS equipped with $M$ antennas and serving $K$ single-antenna users, where $\mathcal{M}=\{1,\ldots,M\}$ and $\mathcal{K}=\{1,\ldots,K\}$. The FSIM contains $L$ layers, where each layer is equipped with $N = N_h \cdot N_v$ meta-atoms, with set as  $\mathcal{L} = \{1, \ldots,L\}$, $\mathcal{N} = \{1, \ldots, N\}$. Let $\mathcal{N}_h = \{1, \ldots,N_h\}$, and $\mathcal{N}_v = \{1, \ldots,N_v\}$ denote the sets of horizontal and vertical meta-atoms, respectively. In the $l$-th layer, the phase-shift of the $n$-th meta-atom can be expressed as $ e^{j\theta_n^{l}}$, where $\theta_n^{l} \in [0, 2\pi)$, and its matrix form is denoted as $\boldsymbol{\Phi}_l = \operatorname{diag}(e^{j\theta_1^{l}}, \ldots, e^{j\theta_N^{l}}) \in \mathbb{C}^{N \times N}.$ The positions of the BS, user $k$, and the $(n_h,n_v)$-th FSIM meta-atom on the $l$-th layer are respectively defined as $\mathbf{x}^{\mathrm{BS}}=[x^{\mathrm{BS}},\,y^{\mathrm{BS}},\,z^{\mathrm{BS}}]^{\mathrm{T}}$, $\mathbf{x}^{\mathrm{UE}}_{k}=[x^{\mathrm{UE}}_{k},\,y^{\mathrm{UE}}_{k},\,z^{\mathrm{UE}}_{k}]^{\mathrm{T}}$, and $\mathbf{x}^{l}_{n_h,n_v}=[x^{l}_{n_h,n_v},\,y^{l}_{n_h,n_v},\,z^{l}_{n_h,n_v}]^{\mathrm{T}}$. The FE position set is defined as
$\mathbf{X}^{\mathrm{FE}}
\triangleq
\left\{
\mathbf{x}^{l}_{n_h,n_v}
| \forall
l\in\mathcal{L},
n_h\in\mathcal{N}_h,
n_v\in\mathcal{N}_v
\right\}$.
In each layer, the meta-atoms are fluidically movable within the layer area, subject to the following constraints
\begingroup
\allowdisplaybreaks
\begin{subequations}\label{eq:FA_mobility}
\begin{align}
& x^{l}_{n_h,n_v}-x^{l}_{n_h-1,n'_v}> d_{x,\mathrm{th}},~ \forall\, n_h\in\mathcal{N}_h,\; n_v,n'_v\in\mathcal{N}_v, \label{eq:FA_mobility_a}\\
& z^{l}_{n_h,n_v}-z^{l}_{n'_h,n_v-1}> d_{z,\mathrm{th}},~ \forall\, n_h,n'_h\in\mathcal{N}_h,\; n_v\in\mathcal{N}_v, \label{eq:FA_mobility_c}\\
& x^{l}_{1,n_v}\ge 0,\; x^{l}_{N_h,n_v}\le L_h,~ \forall\, n_v\in\mathcal{N}_v, \label{eq:FA_mobility_d}\\
& z^{l}_{n_h,1}\ge 0,\; z^{l}_{n_h,N_v}\le L_v,~ \forall\, n_h\in\mathcal{N}_h. \label{eq:FA_mobility_e}\\
& y^{l}_{n_h,n_v}=y^{l}_{n'_h,n'_v}, \notag\\
& \forall\, n_h,n'_h\in\mathcal{N}_h,\; n_v,n'_v\in\mathcal{N}_v,\; n_h\neq n'_h,\; n_v\neq n'_v. \label{eq:FA_mobility_b}
\end{align}
\end{subequations}
\endgroup
Constraints \eqref{eq:FA_mobility_a} and \eqref{eq:FA_mobility_c} ensure that the inter-element spacing of horizontal and vertical axis should be respectively larger than $d_{x,\mathrm{th}}$ and $d_{z,\mathrm{th}}$ to avoid collisions. Moreover, \eqref{eq:FA_mobility_d} and \eqref{eq:FA_mobility_e} guarantee that all meta-atoms are within the predefined
size of each layer. All meta-atoms at layer $l$ share the same y-coordinate in \eqref{eq:FA_mobility_b}, since they are located on the same 2D panel.

	Let $\mathbf{W}_1\in \mathbb{C}^{N \times M}$ denote the channel between BS and first layer, whereas $\mathbf{W}_l \in \mathbb{C}^{N\times N},~\forall\, l\in\mathcal{L}\setminus\{1\}$ denotes the channels between layers $(l-1)$ and $l$ of FSIM. The channel coefficient from the $\tilde{n}$-th meta-atom on layer $(l\!-\!1)$ to the $n$-th meta-atom on layer $l$ is given by \cite{4}
\begin{align}\label{eq:wln}
w^{l}_{n,\tilde{n}}
= \frac{A_t \cos\chi^{l}_{n,\tilde{n}}}{d^{l}_{n,\tilde{n}}}
\left(\frac{1}{2\pi d^{l}_{n,\tilde{n}}} - j\frac{1}{\lambda}\right)
e^{j2\pi d^{l}_{n,\tilde{n}}/\lambda},
\end{align}
where $A_t$ denotes the area of each meta-atom. $\chi_{n,\tilde{n}}^{l}$ and $d^{l}_{n,\tilde{n}}$ indicate propagation direction and distance between the $\tilde{n}$-th meta-atom on layer $(l-1)$ to the $n$-th meta-atom on layer $l$, respectively. The geometric terms are given by
$\cos\chi^{l}_{n,\tilde{n}}=\frac{d_{\mathrm{layer}}}{d^{l}_{n,\tilde{n}}}$
and
$d^{l}_{n,\tilde{n}}=\sqrt{d_{\mathrm{layer}}^{2}+\left(x^{l}_{n_h,n_v}-x^{l-1}_{n_h,n_v}\right)^2 + \left(z^{l}_{n_h,n_v} z^{l-1}_{n_h,n_v}\right)^2}$, where $d_{\mathrm{layer}}$ is the inter-layer spacing. The cascaded near-field channel of FSIM can be expressed as
\begin{align}
\mathbf{S} = \mathbf{\Phi}_L \mathbf{W}_L \cdots \mathbf{\Phi}_2 \mathbf{W}_2 \mathbf{\Phi}_1 \in \mathbb{C}^{N\times N}.
\end{align}
Let $\bar{\mathbf{g}}_{k}$ denote the channel from the last FSIM layer to user $k$, which is given by
\begin{align}\label{eq:gSk}
\bar{\mathbf{g}}_{k}
= \sqrt{\beta_k}\left(
\sqrt{\frac{\kappa_{k}}{\kappa_{k}+1}}\,\mathbf{g}^{\mathrm{LoS}}_{k}
+\sqrt{\frac{1}{\kappa_{k}+1}}\,\mathbf{g}^{\mathrm{NLoS}}_{k}
\right),
\end{align}
where $\beta_k$ represents the large-scale fading, and $\kappa_k$ indicates the Rician factor. Notation $\mathbf{g}^{\mathrm{LoS}}_{k}=\mathrm{vec}\!\left(\mathbf{A}^{\mathrm{FE}}\right)\in\mathbb{C}^{N\times 1}$ denotes the array response of the last FSIM layer, whereas $\mathrm{vec}(\cdot)$ vectorizes the matrix. The $(n_h,n_v)$-th entry of $\mathbf{A}^{\mathrm{FE}}$ is given by $[\mathbf{A}^{\mathrm{FE}}]_{n_h,n_v}=e^{j\mathbf{k}\mathbf{x}^{L}_{n_h,n_v}}$, associated with $\mathbf{k}=\frac{2\pi}{\lambda}\big[\cos(\theta)\sin(\psi),\,\sin(\theta)\sin(\psi),\,\cos(\psi)\big]$ as the wave number and $\{\theta, \psi\}$ respectively as the azimuth and elevation angle of departure (AoD). Moreover, $\mathbf{g}^{\mathrm{NLoS}}_{k}\sim\mathcal{CN}\!\left(\mathbf{0},\mathbf{R}\right)\in\mathbb{C}^{N\times 1}$, where $\mathbf{R}\in\mathbb{C}^{N\times N}$ is the spatial correlation \cite{16} with the $(n,\tilde{n})$-th FSIM entry given by $[\mathbf{R}]_{n,\tilde{n}}=\mathrm{sinc}\!\left(2d_{n,\tilde{n}}/\lambda\right)$. The effective total channel is given by $\mathbf{g}_k=\bar{\mathbf{g}}_{k}^{\mathrm{H}}\mathbf{S}\mathbf{W}_1\in\mathbb{C}^{1\times M}$. Therefore, the received signal for user $k$ is represented by
$y_k = \mathbf{g}_k\mathbf{p}_k s_k + \mathbf{g}_k\sum_{k'\in\mathcal{K}\setminus\{k\}} \mathbf{p}_{k'} s_{k'} + n_k,$
where $s_k$ and $\mathbf{p}_k\in\mathbb{C}^{M\times 1}$ denote the transmitted data symbol and the beamforming vector, respectively. Notation of $n_k\sim\mathcal{CN}(0,\sigma_k^2)$ denotes the additive white Gaussian noise (AWGN) associated with power $\sigma_k^2$. Therefore, the signal-to-interference-plus-noise ratio (SINR) of user $k$ is attained as
\begin{equation}\label{eq:SINR}
\gamma_k=\frac{\left|\mathbf{g}_k\mathbf{p}_k\right|^2}{\sum_{k'\in\mathcal{K}\setminus\{k\}}\left|\mathbf{g}_k\mathbf{p}_{k'}\right|^2+\sigma_k^{2}}.
\end{equation}
Then the achievable rate of user $k$ is
$R_k=\log_2\!\left(1+\gamma_k\right).$

The objective is to maximize the sum-rate by jointly optimizing the transmit beamforming $\mathbf{p}_{k}$, the phase-shifts $\mathbf{\Phi}_{l}$, and the element positions of FSIM $\mathbf{X}^{\mathrm{FE}}$. The problem can be formulated as 
\begingroup
\allowdisplaybreaks
\begin{subequations}\label{eq:prob_sumrate}
\begin{align}
\max_{\mathbf{p}_k,\,\boldsymbol{\Phi}_l,\,\mathbf{X}^{\mathrm{FE}}}\quad 
& \sum_{k\in\mathcal{K}} R_k \label{eq:prob_sumrate_obj}\\
\text{s.t.}\quad 
& \sum_{k\in\mathcal{K}}\|\mathbf{p}_k\|^2 \le P_{\max}, \label{eq:prob_sumrate_p}\\
& \theta_n^{l}\in[0,2\pi),\ \forall\, n\in\mathcal{N},\ \forall\, l\in\mathcal{L}, \label{eq:prob_sumrate_theta}\\
& \mathbf{X}^{\mathrm{FE}}\in\mathcal{R}_{\mathbf{X}}. \label{eq:prob_sumrate_pos}
\end{align}
\end{subequations}
\endgroup
Constraint \eqref{eq:prob_sumrate_p} guarantees maximum transmit power at the BS as $P_{\max}$, \eqref{eq:prob_sumrate_theta} limits phase-shift for each meta-atoms, and \eqref{eq:prob_sumrate_pos} restricts the deployed boundary within each FSIM layer, where $\mathcal{R}_{\mathbf{X}}=\{\eqref{eq:FA_mobility_a}\text{--}\eqref{eq:FA_mobility_b}\}$. Note that the problem is challenging due to the coupling configuration variables. Therefore, we develop solution to optimize the respective subproblems.

\section{Proposed Solution}
\label{sec_alg}

\subsection{Beamforming Optimization}

Assume FE positions  $\mathbf X^{\mathrm{FE}}$ and phase-shifts $\mathbf{\Phi}_{l}$ are fixed, and we optimize the beamforming $\mathbf{P} = [\mathbf{p}_1, \ldots, \mathbf{p}_k]$. Then the subproblem can be formulated as
\begin{equation}\label{eq:subprob_P}
\max_{\mathbf P}\ \sum_{k\in\mathcal K} R_k(\mathbf P)
\ \ \text{s.t.}\ \eqref{eq:prob_sumrate_p}.
\end{equation}
We transform \eqref{eq:subprob_P} by introducing an auxiliary variable $\mathbf{F}_k \triangleq \mathbf{p}_k \mathbf{p}_k^{\mathrm{H}}$ with the constraint of $\mathrm{rank}(\mathbf{F}_k)=1$. The objective function can be rewritten as
$R_k(\mathbf F)= \log_2\!\left(1+\frac{\operatorname{Tr}\left(\mathbf F_k\mathbf g_k^{\mathrm{H}}\mathbf g_k\right)}{\sum_{j\in\mathcal K\setminus\{k\}}\operatorname{Tr}\left(\mathbf F_j\mathbf g_k^{\mathrm{H}}\mathbf g_k\right)+\sigma_k^{2}}\right).$ Therefore, the subproblem \eqref{eq:subprob_P} can be attained as
\begin{subequations}\label{eq:F}
\begin{align}
\max_{\mathbf F}\quad & \sum_{k\in\mathcal K} R_k(\mathbf F) \label{eq:11a}\\
\text{s.t.}\quad 
& \mathbf F_k \succeq 0,\ \forall k\in\mathcal K, \quad 
\sum_{k\in\mathcal K}\operatorname{Tr}(\mathbf F_k)\le P_{\max}, \label{eq:11b}\\
& \operatorname{rank}(\mathbf F_k)=1,\ \forall k\in\mathcal K. \label{eq:11c}
\end{align}
\end{subequations}
Note that \eqref{eq:11a} and \eqref{eq:11c} are non-convex. We first reformulate the objective in \eqref{eq:11a} by writing
$R_k(\mathbf F)= f_k(\mathbf F)- z_k(\mathbf F)$, where
$f_k(\mathbf F)=\log_2\left(\sum_{j\in\mathcal K}\operatorname{Tr}(\mathbf F_j\mathbf g_k^{\mathrm{H}}\mathbf g_k)+\sigma_k^{2}\right)$
and $z_k(\mathbf F)=\log_2\left(\sum_{j\in\mathcal K\setminus\{k\}}\operatorname{Tr}(\mathbf F_j\mathbf g_k^{\mathrm{H}}\mathbf g_k)+\sigma_k^{2}\right)$. However, $z_k(\mathbf F)$ is still non-convex. To tackle this issue, we adopt the SCA with the first-order Taylor approximation as $\hat{z}_k(\mathbf F)\approx z_k(\mathbf F^{(t)})+\operatorname{Tr}\big(\nabla z_k^{\mathrm{T}}(\mathbf F^{(t)})(\mathbf F-\mathbf F^{(t)})\big)$, where $\nabla z_k^{\mathrm{T}}(\mathbf F^{(t)})=\dfrac{\mathbf g_k^{\mathrm{H}}\mathbf g_k}{\left(\sum_{j\in\mathcal K\setminus\{k\}}\left(\operatorname{Tr}(\mathbf F_j\,\mathbf g_k^{\mathrm{H}}\mathbf g_k)+\sigma_k^{2}\right)\right)\cdot\ln 2}$. To handle the rank-one constraint in \eqref{eq:11c}, we transform it into a penalty term. When $\mathbf F_k$ is rank-one, it satisfies
\begin{align}\label{eq:rank1_penalty}
\|\mathbf F_k\|_{*}-\|\mathbf F_k\|_{2}=0,\quad \forall k\in\mathcal K,
\end{align}
where $\|\mathbf F_k\|_{*}=\sum_{i}\nu_{k,i}$ denotes the nuclear norm, $\|\mathbf F_k\|_{2}=\nu_{k,1}$ denotes the spectral norm, and $\nu_{k,i}$ indicates the $i$-th largest eigen value of $\mathbf F_k$. Since \eqref{eq:rank1_penalty} is non-convex due to the spectral-norm $\|\mathbf F_k\|_2$, we apply the first-order Taylor approximation to obtain a convex upper bound as
\begin{align}\label{eq:taylor_specnorm}
\|\mathbf{F}_k\|_{*}-\|\mathbf{F}_k\|_{2} \leq \|\mathbf{F}_k\|_{*}-\|\bar{\mathbf{F}}_k\|_{2}, \quad \forall k\in\mathcal{K},
\end{align}
where $\|\bar{\mathbf F}_k\|_{2}
= \|\mathbf F_k^{(t)}\|_{2}
+ \operatorname{Tr}\Big(\mathbf e_k^{(t)}\big(\mathbf e_k^{(t)}\big)^{\mathrm{H}}
\big(\mathbf F_k-\mathbf F_k^{(t)}\big)\Big)$, and $\mathbf e_k^{(t)}$ is the eigenvector corresponding to the largest eigenvalue of $\mathbf F_k^{(t)}$ in the $t$-th iteration. Therefore, problem \eqref{eq:F} can be approximated as
\begingroup
\allowdisplaybreaks
\begin{subequations}\label{eq:maxF}
\begin{align}
\max_{\mathbf F}\quad 
& \sum_{k\in\mathcal K}\Big[ f_k(\mathbf F)-\hat z_k(\mathbf F)
-\frac{1}{\mu_1}\big(\|\mathbf F_k\|_{*}-\|\bar{\mathbf F}_k\|_{2}\big)\Big]\label{eq:14a}\\
\text{s.t.}\quad 
& \eqref{eq:11b}, \label{eq:14b}
\end{align}
\end{subequations}
\endgroup
where $\mu_1$ denotes the penalty factor. We initialize the penalty factor $\mu_1$ with a large value and update it as $\mu_1 \leftarrow \alpha_1\cdot\mu_1$, where $0<\alpha_1<1$, until the rank-one penalty satisfies 
$\max\big\{\|\mathbf F_k\|_{*}-\|\bar{\mathbf F}_k\|_{2}\big\}\le \varepsilon$. The subproblem \eqref{eq:maxF} becomes convex and can be solved by convex optimization tools. 

\subsection{Phase-Shift Optimization}

First, we expand $\bigl| \mathbf{g}_k \mathbf{p}_k \bigr|^2 \allowbreak
= \allowbreak \Bigl| \boldsymbol{\psi}_{l}^{\mathrm{H}}\!\bigl(\mathrm{diag}(\bar{\mathbf{g}}_{k}^{\mathrm{H}}) \mathbf{A}_l \mathbf{C}_l \mathbf{W}_1 \mathbf{p}_k \bigr) \Bigr|^2 \allowbreak
= \allowbreak \bigl| \boldsymbol{\psi}_{l}^{\mathrm{H}}\mathbf{q}_{k,k} \bigr|^2 \allowbreak
= \allowbreak \boldsymbol{\psi}_{l}^{\mathrm{H}}\mathbf{Q}_{k,k}\boldsymbol{\psi}_{l}$,
where $\boldsymbol{\psi}_{l} = [e^{j\theta_1^{l}},\ldots,e^{j\theta_N^{l}}]^{\mathrm{T}}$,
$\mathbf{Q}_{k,k} = \mathbf{q}_{k,k}\mathbf{q}_{k,k}^{\mathrm{H}}$, 
$\mathbf{q}_{k,k} = \mathrm{diag}(\bar{\mathbf{g}}_{k}^{\mathrm{H}}) \mathbf{A}_l \mathbf{C}_l \mathbf{W}_1 \mathbf{p}_k$,
$\mathbf{A}_l = \mathbf{\Phi}_L \mathbf{W}_L \cdots \mathbf{\Phi}_{l+1}\mathbf{W}_{l+1}$, 
and $\mathbf{C}_l = \mathbf{W}_l \mathbf{\Phi}_{l-1}\mathbf{W}_{l-1}\cdots \mathbf{\Phi}_1$. We define $\mathbf{V}_{l}=\boldsymbol{\psi}_{l}\boldsymbol{\psi}_{l}^{\mathrm{H}}$, and the objective can be reformulated as
$R_k(\mathbf{V}_l)=\log_2\!\left(1+\dfrac{\mathrm{Tr}\!\left(\mathbf{V}_l\mathbf{Q}_{k,k}\right)}
{\sum_{j\in\mathcal{K}\setminus\{k\}}\mathrm{Tr}\!\left(\mathbf{V}_l\mathbf{Q}_{k,j}\right)+\sigma_k^2}\right)$. Therefore, the subproblem can be written as
\begin{subequations}\label{eq:V_problem}
\begin{align}
\max_{\mathbf{V}_l} \quad 
& \sum_{k\in\mathcal{K}} R_k(\mathbf{V}_l) \label{eq:V_problem_a}\\
\text{s.t.}\quad
& \mathrm{diag}(\mathbf{V}_l) = 1,
\mathbf{V}_l \succeq 0,
\mathrm{rank}(\mathbf{V}_l) = 1,\ \forall l \in \mathcal{L}. 
\label{eq:V_problem_b}
\end{align}
\end{subequations}
We rewrite the objective function in \eqref{eq:V_problem} into a form
$R_k(\mathbf{V}_l)= f_k(\mathbf{V}_l)-z_k(\mathbf{V}_l)$, where 
$f_k(\mathbf{V}_l)=\log_2\!\Big(\sum_{j\in\mathcal{K}}\mathrm{Tr}\big(\mathbf{V}_l\mathbf{Q}_{k,j}\big)+\sigma_k^2\Big)$ and 
$z_k(\mathbf{V}_l)=\log_2\!\Big(\sum_{j\in\mathcal{K}\setminus\{k\}}\mathrm{Tr}\big(\mathbf{V}_l\mathbf{Q}_{k,j}\big)+\sigma_k^2\Big)$. Since $z_k(\mathbf{V}_l)$ is non-convex, we adopt the first-order Taylor approximation as
$\hat{z}_k(\mathbf{V}_l)\approx z_k\!\left(\mathbf{V}_l^{(t')}\right)+\mathrm{Tr}\!\left(\nabla z_k^{\mathrm{T}}\!(\mathbf{V}_l^{(t')})(\mathbf{V}_l-\mathbf{V}_l^{(t')})\right)$,
where
$\nabla z_k\!\left(\mathbf{V}_l^{(t')}\right)=\dfrac{\sum_{j\in\mathcal{K}\setminus\{k\}}\mathbf{Q}_{k,j}}{\left(\sum_{j\in\mathcal{K}\setminus\{k\}}\mathrm{Tr}\!\left(\mathbf{V}_l\mathbf{Q}_{k,j}\right)+\sigma_k^{2}\right)\ln 2}$. The rank-one constraint can be equivalently expressed as
\begin{align}\label{eq:rank1_nuc_spec}
\|\mathbf{V}_l\|_* - \|\mathbf{V}_l\|_2 = 0,\ \forall\, l\in\mathcal{L},
\end{align}
where $\|\mathbf{V}_l\|_*=\sum_{i}\varsigma_{l,i}$ and $\|\mathbf{V}_l\|_2=\varsigma_{l,1}$. Note that $\varsigma_{l,i}$ denotes the $i$-th largest eigenvalue of $\mathbf{V}_l$.
By applying the first-order Taylor approximation to \eqref{eq:rank1_nuc_spec}, we can obtain
\begin{align}\label{eq:18_taylor_ineq}
\|\mathbf{V}_l\|_*-\|\mathbf{V}_l\|_2 \le \|\mathbf{V}_l\|_*-\|\bar{\mathbf{V}}_l\|_2,\ \forall\, l\in\mathcal{L},
\end{align}
where
$\|\bar{\mathbf{V}}_l\|_2=\|\mathbf{V}_l^{(t)}\|_2+\mathrm{Tr}\Big(\mathbf{q}_l^{(t)}\big(\mathbf{q}_l^{(t)}\big)^{\mathrm{H}} \big(\mathbf{V}_l-\mathbf{V}_l^{(t)}\big)\Big)$,
and $\mathbf{q}_l^{(t)}$ is the eigenvector corresponding to the largest eigenvalue of $\mathbf{V}_l^{(t)}$ at the $t$-th iteration.
Therefore, \eqref{eq:V_problem} can be reformulated as
\begin{subequations}\label{eq:subprob_Vl}
\begin{align}
\max_{\mathbf{V}_l}\quad 
& \sum_{k\in\mathcal{K}}\Big[f_k(\mathbf{V}_l)-\hat{z}_k(\mathbf{V}_l)
-\frac{1}{\mu_2}\big(\|\mathbf{V}_l\|_*-\|\bar{\mathbf{V}}_l\|_2\big)\Big] \label{eq:subprob_Vl_a}\\
\text{s.t.}\quad 
& \mathrm{diag}(\mathbf{V}_l)=1, \  \mathbf{V}_l\succeq 0, \ \forall l\in\mathcal{L}, \label{eq:subprob_Vl_b}
\end{align}
\end{subequations}
where $\mu_2$ denotes the penalty factor. It is initialized as a large value and updated as $\mu_2 \leftarrow \alpha_2\cdot\mu_2$, where $0<\alpha_2<1$, until the rank-one penalty satisfies
$\max_{l}\left\{\|\mathbf{V}_{l}\|_{*}-\|\bar{\mathbf{V}}_{l}\|_{2}\right\}\leq \varepsilon$. The subproblem \eqref{eq:subprob_Vl} becomes convex and can be efficiently solved by any convex optimization tools.

\subsection{FE Position Optimization}

Here, we optimize the positions of FEs. The position optimization subproblem is formulated as
\begin{align}\label{eq:pos_opt_problem}
\max_{\mathbf{x},\mathbf{z}} \quad
\sum_{k\in\mathcal{K}} R_k(\mathbf{x},\mathbf{z}),\quad
\text{s.t.}\quad
\eqref{eq:FA_mobility_a}-\eqref{eq:FA_mobility_b},
\end{align}
where $\mathbf{x}\triangleq\{x_{n_h,n_v}^{l}| l\in\mathcal{L}, n_h\in\mathcal{N}_h, n_v\in\mathcal{N}_v\}$ and 
$\mathbf{z}\triangleq\{z_{n_h,n_v}^{l}| l\in\mathcal{L}, n_h\in\mathcal{N}_h, n_v\in\mathcal{N}_v\}$ denote the horizontal and vertical coordinates of FEs, respectively. To solve this problem, a projected gradient-based algorithm is adopted to iteratively optimize the FE positions using the corresponding position gradients. The gradient of $R_k(\mathbf{x},\mathbf{z})$ is given by
\begingroup
\footnotesize
\allowdisplaybreaks
\begin{align}
\nabla_{\boldsymbol{\xi}} R_k(\mathbf{x},\mathbf{z})
&=
\frac{1}{\ln 2}
\Bigg(
\frac{\sum_{i\in\mathcal{K}}
\nabla_{\boldsymbol{\xi}} J_{k,i}(\boldsymbol{\xi})}
{\sum_{i\in\mathcal{K}}
J_{k,i}(\boldsymbol{\xi})+\sigma_k^2}
-
\frac{\sum_{i\in\mathcal{K}\setminus\{k\}}
\nabla_{\boldsymbol{\xi}} J_{k,i}(\boldsymbol{\xi})}
{\sum_{i\in\mathcal{K}\setminus\{k\}}
J_{k,i}(\boldsymbol{\xi})+\sigma_k^2}
\Bigg),
\label{eq:grad_Rk}
\end{align}
\endgroup
where $\boldsymbol{\xi}\in\{\mathbf{x},\mathbf{z}\}$ and $J_{k,i}(\mathbf{x},\mathbf{z})
\triangleq
\left|\mathbf{g}_k(\mathbf{x},\mathbf{z})\mathbf{p}_i\right|^2$. Each element of $\nabla_{\boldsymbol{\xi}} J_{k,i}(\boldsymbol{\xi})$ can be expressed as
\begingroup
\allowdisplaybreaks
\begin{align} \label{eq:grad_Jki}
\nabla_{\boldsymbol{\xi}} J_{k,i}(\boldsymbol{\xi})
&=
\Bigg[
\frac{\partial J_{k,i}(\boldsymbol{\xi})}{\partial \xi_{1,1}^{1}},
\frac{\partial J_{k,i}(\boldsymbol{\xi})}{\partial \xi_{1,2}^{1}},
\ldots,
\frac{\partial J_{k,i}(\boldsymbol{\xi})}{\partial \xi_{N_h,N_v}^{L}}
\Bigg]^{\mathrm{T}}.
\end{align}
\endgroup
For intermediate layers of FSIM, i.e., $l \neq L$, the partial derivative can be obtained as
\begingroup
\footnotesize
\allowdisplaybreaks
\begin{align}
& \frac{\partial J_{k,i}(\boldsymbol{\xi})}
{\partial \xi_{n_h,n_v}^{l}}
= 2\Re \Bigg\{
\Bigg[
\bar{\mathbf{g}}_{k}
(
\prod_{\substack{u=L\\u\leftarrow u-1}}^{l+2}
\mathbf{\Phi}_u \mathbf{W}_u
)
\mathbf{\Phi}_{l+1}
\cdot
\left(
\frac{\partial \mathbf{W}_{l+1}}
{\partial \xi_{n_h,n_v}^{l}}
\mathbf{\Phi}_{l}\mathbf{W}_{l} \right.
\notag\\
&\left. +
\mathbf{W}_{l+1}\mathbf{\Phi}_{l}
\frac{\partial \mathbf{W}_{l}}
{\partial \xi_{n_h,n_v}^{l}}
\right)
 \cdot
(
\prod_{\substack{u=l-1\\u\leftarrow u-1}}^{1}
\mathbf{\Phi}_{u}\mathbf{W}_{u}
)
\mathbf{p}_{i}
\Bigg]^{*}
\mathbf{g}_{k}\mathbf{p}_{i}
\Bigg\},  \  \xi \in \{x,z\},
\label{eq:dJ_dxi_l_neq_L}
\end{align}
\endgroup
where $(\cdot)^*$ is the conjugate operation.
The derivatives of $\mathbf{W}_l$ and $\mathbf{W}_{l+1}$ in \eqref{eq:dJ_dxi_l_neq_L} are given by
\begingroup
\footnotesize
\allowdisplaybreaks
\begin{align} \label{eq:dW_unified}
\frac{\partial \mathbf{W}_{l+\delta}}
{\partial \xi_{n_h,n_v}^{\,l}}
&=
\frac{
\xi_{n_h,n_v}^{\,l}
-
\xi_{\tilde n_h,\tilde n_v}^{\,l+2\delta-1}
}
{d_{n,\tilde n}^{\,l}}
\Bigg[
-\frac{2bcA_t d_{\mathrm{layer}}}
{(d_{n,\tilde n}^{\,l})^{3}}
-\frac{ac}
{2\pi (d_{n,\tilde n}^{\,l})^{2}}
\notag \\
&
+
ab\left(
j\frac{2\pi}{\lambda}
e^{j\frac{2\pi d_{n,\tilde n}^{\,l}}{\lambda}}
\right)
\Bigg], \ \delta \in\{0,1\},
\end{align}
\endgroup
where $a=\frac{A_t\cos x_{n,\tilde{n}}^{l}}{d_{n,\tilde{n}}^{l}}$, $b=\frac{1}{2\pi d_{n,\tilde{n}}^{l}}-j\frac{1}{\lambda}$, and $c=e^{j2\pi d_{n\tilde{n}}^{l}/\lambda}$.
For \( l = L \), the partial derivative $J_{k,i}(\boldsymbol{\xi})$can be expressed as
\begingroup
\footnotesize
\allowdisplaybreaks
\begin{align} \label{eq:dJ_dxi_L}
\frac{\partial J_{k,i}(\boldsymbol{\xi})}{\partial \xi_{n_h,n_v}^{L}}
& =
2\Re\Bigg\{ 
\Bigg[
\frac{\partial \bar{\mathbf{g}}_{k}}{\partial \xi_{n_h,n_v}^{L}}
\mathbf{\Phi}_{L}\mathbf{W}_{L}
+
\mathbf{g}_{S,k}\mathbf{\Phi}_{L}
\frac{\partial \mathbf{W}_{L}}{\partial \xi_{n_h,n_v}^{L}}
\Bigg]
\notag \\
& \cdot
(
\prod_{\substack{u=L-1\\u\leftarrow u-1}}^{1}\mathbf{\Phi}_{u}\mathbf{W}_{u}
)
\mathbf{p}_{i}
\Bigg]^{*}
\mathbf{g}_{k}\mathbf{p}_{i}
\Bigg\}, \ \xi \in \{x,z\}.
\end{align}
\endgroup
When $l=L$, the derivative of $\mathbf{W}_{L}$ can be directly obtained from \eqref{eq:dW_unified} by setting $l=L$ and $\delta=0$, and the gradient of $\bar{\mathbf{g}}_{k}$ in \eqref{eq:dJ_dxi_L} is given by
\begingroup
\footnotesize
\allowdisplaybreaks
\begin{align}\label{eq:dgsk_dxi}
&\frac{\partial \bar{\mathbf{g}}_{k}}
{\partial \xi_{n_h,n_v}^{L}}
=
\sqrt{\beta_k}
\sqrt{\frac{\kappa_{k}}{\kappa_{k}+1}}
\cdot
j\frac{2\pi}{\lambda}
k_{\xi}
\exp\Bigg[
j\frac{2\pi}{\lambda}
\Big(
x_{n_h,n_v}^{L}\cos(\theta)
\notag \\
&\quad\  \cdot \sin(\psi)
+
y_{n_h,n_v}^{L}\sin(\theta)\sin(\psi)
+
z_{n_h,n_v}^{L}\cos(\psi)
\Big)
\Bigg],
\end{align}
\endgroup
where $k_x = \cos(\theta)\sin(\psi)$ and $k_z = \cos(\psi)$.
Therefore, the gradient of the sum-rate with respect to $\boldsymbol{\xi}$ can be expressed as 
$\nabla_{\boldsymbol{\xi}} R_{\mathrm{sum}}(\mathbf{x},\mathbf{z})
= \sum_{k\in\mathcal{K}} \nabla_{\boldsymbol{\xi}} R_k(\mathbf{x},\mathbf{z})$. 
Then, the solution is iteratively updated as
\begin{align}\label{eq:FA_position_update}
\boldsymbol{\xi}^{(t)}
\leftarrow
\Pi_{\mathbf{X}^{\mathrm{FE}}}
\left(
\boldsymbol{\xi}^{(t-1)}
+
\eta
\nabla_{\boldsymbol{\xi}}
R_{\mathrm{sum}}(\mathbf{x},\mathbf{z})
\right), \boldsymbol{\xi}\in\{\mathbf{x},\mathbf{z}\},
\end{align}
where
$\Pi_{\mathbf{X}^{\mathrm{FE}}}
=\min\!\left(L_{\xi},\max\!\left(\chi_{n_h,n_v}^{\mathrm{T}},0\right)\right)$, 
with $\chi_{n_h,n_v}^{\mathrm{T}}
=\max\!\left(\xi_{n_h,n_v}^{\mathrm{T}},
\xi_{n_h',n_v'}^{\mathrm{T}}+d_{\xi,\mathrm{th}}\right)$ and $\xi\in\{x,z\}$, indicates the projection function mapping. The overall computational complexity of the proposed AO-based algorithm elaborated in Algorithm \ref{alg:proposed_AO} is dominated by the beamforming in \eqref{eq:maxF}, the phase-shift in \eqref{eq:subprob_Vl}, and the position calculation in \eqref{eq:FA_position_update}, which can be shown to be $
\mathcal{O}\left[
T_{\mathrm{AO}}
\left(
KM^{3.5}
+
LN^{3.5}
+
K^2LN
\right)
\right],
$
where $T_{\mathrm{AO}}$ denotes the number of AO iterations. The three terms correspond to the computational complexities of the beamforming optimization, FSIM configuration optimization, and FE position optimization, respectively.

\begin{algorithm}[!t]
\footnotesize
\caption{Proposed AO-based Algorithm}
\label{alg:proposed_AO}
{
\textbf{Input:} 
$\mathbf{P}^{(0)}$, $\boldsymbol{\Phi}^{(0)}$, 
$\mathbf{X}^{\mathrm{FE},(0)}$, $\epsilon \geq 0$, 
$T_{\max}$, $t=1$

\Repeat{
$ t>T_{\max} \ \mathrm{ or }\ 
\bigl| R_{\mathrm{sum}}(
\mathbf{P}^{(t)},
\boldsymbol{\Phi}^{(t)},
\mathbf{X}^{\mathrm{FE},(t)}
)
-
R_{\mathrm{sum}}(
\mathbf{P}^{(t\!-\!1)},
\boldsymbol{\Phi}^{(t\!-\!1)},
\mathbf{X}^{\mathrm{FE},(t\!-\!1)}
)
\bigr|
\leq \epsilon
$
}{
Update $\mathbf{P}^{(t)}$ by solving \eqref{eq:maxF} with fixed 
$\boldsymbol{\Phi}^{(t-1)}$ and 
$\mathbf{X}^{\mathrm{FE},(t-1)}$

Update $\boldsymbol{\Phi}^{(t)}$ by solving \eqref{eq:subprob_Vl} with fixed 
$\mathbf{P}^{(t)}$ and 
$\mathbf{X}^{\mathrm{FE},(t-1)}$

Update $\mathbf{X}^{\mathrm{FE},(t)}$ using \eqref{eq:FA_position_update} 
and project it onto the feasible set

$t \leftarrow t+1$
}

\textbf{Output:} 
Optimized $\mathbf{P}^{\star}=\mathbf{P}^{(t)}$, 
$\boldsymbol{\Phi}^{\star}=\boldsymbol{\Phi}^{(t)}$, 
$\mathbf{X}^{\mathrm{FE},\star}=\mathbf{X}^{\mathrm{FE},(t)}$
}
\end{algorithm}

\section{Simulation Results} \label{sec_sim}

Simulation results are presented to evaluate the performance of the proposed AO-based algorithm in an FSIM-aided network. The pertinent parameters are set as follows. The carrier frequency is set to $28$ GHz. The BS is equipped with $M=4$ antennas and the corresponding inter-antenna spacing is $\lambda/2$, which serves $K=4$ single-antenna users. The BS is located at $(0,0,10)$ m, whereas its antennas are uniformly deployed along the $x$-axis. The FSIM is located at $(0, 0.01, 10)$ m. The FSIM consists of $L=3$ layers, associated with each layer having $N_h=6$ horizontal and $N_v=6$ vertical FEs. The size of each layer is $0.2~\mathrm{m}\times0.2~\mathrm{m}$. The total thickness of the FSIM is set to $5\lambda$, and the inter-layer distance is $d_{\mathrm{layer}}=5\lambda/L$. The minimum spacing between adjacent elements is set to $\lambda/2$. The users are randomly distributed within a circular region with radius $50$ m, whose center is located $100$ m away from the last FSIM layer. The large-scale channel power gain is modeled as $\beta_k=C_0(d_k/d_0)^{-\varpi}$, where $C_0 = -30$ dB, $d_0=1$ m, and $\varpi=2.5$. The Rician factor is set to $\kappa_{k}=10$. The maximum transmit power and noise power are set to $P_{\max}=30$ dBm and $\sigma_k^2=-104$ dBm, respectively.

\begin{figure*}[!t]
\centering

\begin{minipage}{1.6in}
\centering
\includegraphics[width=1.6in]{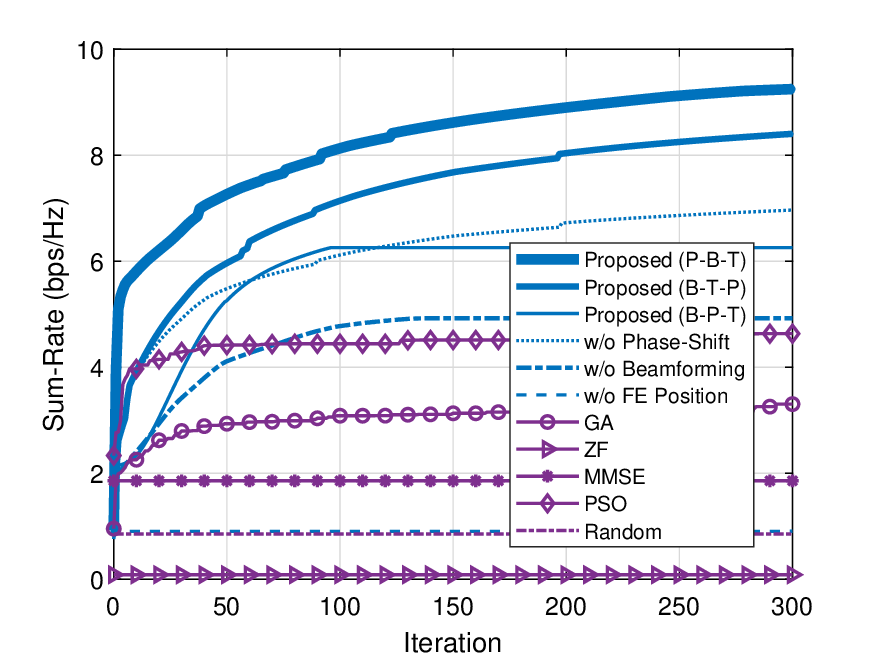}
\caption{Convergence.}
\label{fig:compare_convergence}
\end{minipage}
\hfil
\begin{minipage}{1.6in}
\centering
\includegraphics[width=1.6in]{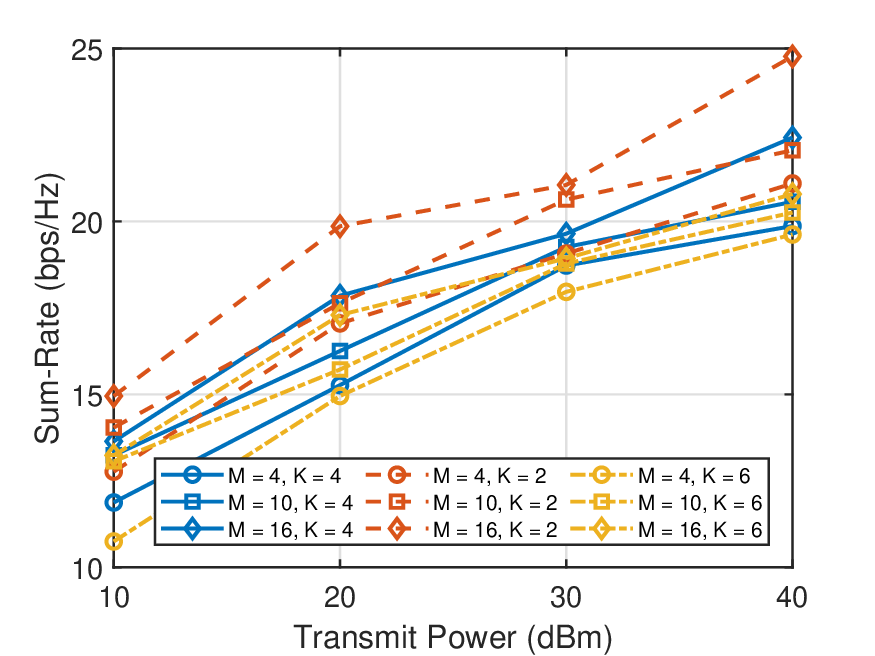}
\caption{Different transmit power.}
\label{fig:power_ant}
\end{minipage}
\hfil
\begin{minipage}{1.6in}
\centering
\includegraphics[width=1.6in]{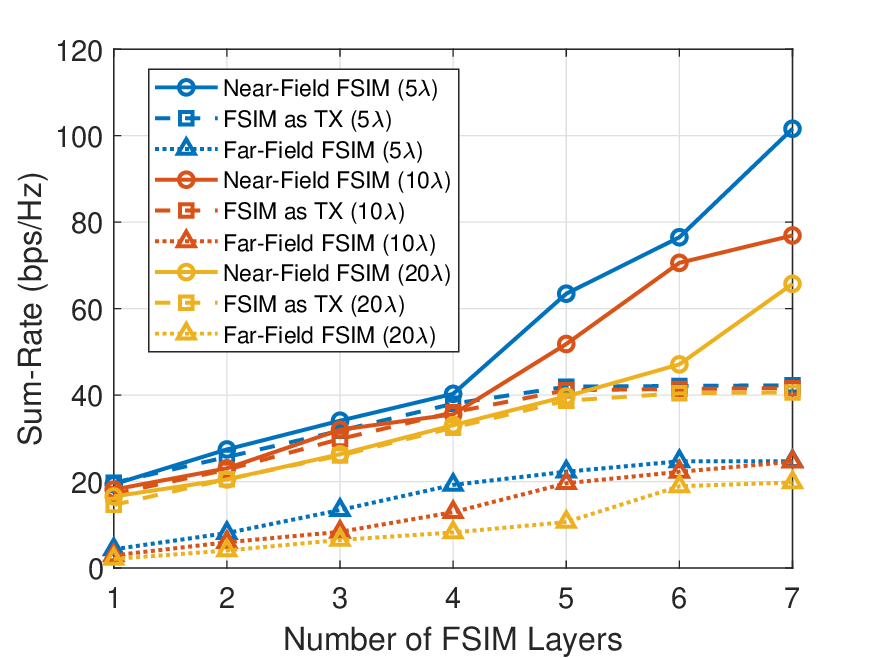}
\caption{Various FSIM characters.}
\label{fig:sumrate_layers_thickness}
\end{minipage}
\hfil
\begin{minipage}{1.6in}
\centering
\includegraphics[width=1.6in]{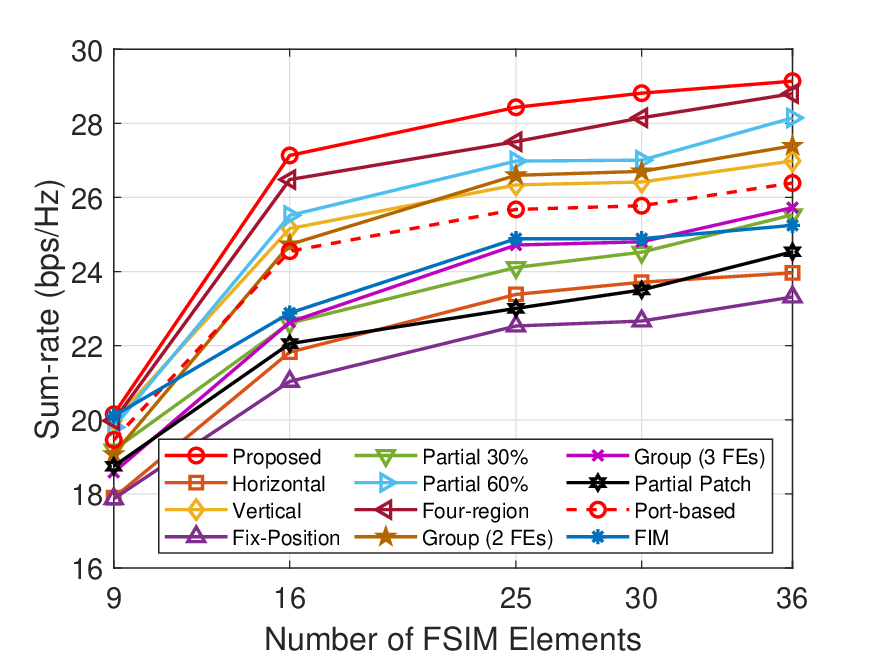}
\caption{Benchmark comparison.}
\label{fig:position}
\end{minipage}

\end{figure*}

Fig.~\ref{fig:compare_convergence} shows the convergence of the proposed solution compared to different benchmarks. For clarity, 'P', 'B', and 'T' denote the position, beamforming, and phase-shift, respectively. Note that metaheuristics of genetic algorithm (GA) \cite{17} and particle swarm optimization (PSO) \cite{18} as well as conventional beamforming of zero-forcing (ZF) and minimum mean square error (MMSE) are compared. It can be observed that the proposed algorithm gradually converges, which verifies the effectiveness of joint optimization of beamforming phase-shift and FE positions. Among the update orders, P-B-T achieves the highest sum-rate since position optimization can provide a more favorable cascaded channel fixed for the beamforming and phase-shift optimization. In contrast, omitting any one of the optimization variables leads to a decrease in sum-rate, which confirms the necessity of full optimization. Moreover, ZF performs a worse rate than that of the random baseline because the effective channels may be highly correlated or interfered. Furthermore, compared to other benchmark algorithms, the proposed AO-based scheme achieves higher sum-rate performance, as heuristic methods potentially provide the local optimal solutions.

Fig.~\ref{fig:power_ant} illustrates the sum-rate performance versus the transmit power for different numbers of BS antennas and users. The sum-rate improves among all settings as the transmit power increases. For a fixed number of users, a larger number of BS antennas achieves a higher sum-rate thanks to higher spatial degrees of freedom. Moreover, the case with $K=2$ outperforms those with $K=4$ and $K=6$, since fewer users introduce less inter-user interference and allow more efficient resource allocation. These results confirm that the sum-rate can be improved by increasing the number of BS antennas under fewer users served.

Fig.~\ref{fig:sumrate_layers_thickness} illustrates the sum-rate performance versus the number of layers under different propagation models and thickness. It can be known that the near-field case achieves the highest sum-rate among all cases due to the finer resolution of beamforming. When the number of layers increases, the sum-rate of the near-field model improves significantly because more layers provide finer-grained spatial degrees of freedom. In contrast, the FSIM operating as a transmitter (FSIM as TX), where the active beamforming matrix $\mathbf{P}$ is omitted, achieves only a moderate rate improvement due to the lack of concentrated beamforming toward the users. Furthermore, the far-field case where FSIM is deployed far away from the BS has the lowest sum-rate, since the path loss potentially dominates the beamforming gains at the BS and FSIM. It is also seen that increasing the thickness of FSIM decreases the sum-rate. This is due to the reason that larger thickness will weaken the inter-layer coupling and introduce additional near-field propagation attenuation, which reduces the effective signal power. Therefore, a compact FSIM structure with appropriate thickness can provide high rate performance.

Fig.~\ref{fig:position} shows the effects of different numbers of elements under different fluid array architectures. It can be observed that the sum-rate of all schemes increases with more elements, since more elements provide higher spatial degrees of freedom. Among all benchmarks, the proposed whole surface boundary achieves the highest sum-rate because all elements are allowed to move over the entire region. The cases where FEs move within four regions and where 60$\%$ of FEs are movable achieve rate close to that of the whole boundary case, whereas the other partially-fluid, element-grouping, and patch-boundary cases exhibit degraded rates due to the reduced spatial degrees of freedom of FEs. In contrast, the horizontal- and vertical-only fluid cases provide much lower rate compared to the whole boundary owing to the restricted movement to a single axis. Discrete port-based element selection also has a lower rate. The conventional structure with fixed position and equal inter-element spacing has the worst rate performance because it cannot exploit any spatial reconfigurability of FSIM elements.

\section{Conclusions} \label{sec_con}

In this paper, we have conceived an FSIM-aided communication system, where the meta-atoms of FSIM are allowed to move within predefined regions within FSIM. To maximize the system sum-rate, the BS transmit beamforming, FSIM phase-shifts, and element positions are jointly optimized. An AO-based algorithm is developed to solve the non-convex problem, where the beamforming and phase-shift subproblems are handled by SCA and rank-one relaxation associated with nuclear and spectral norm approximation. The subproblem of FE positions is solved by using a projected gradient-based scheme. Simulation results demonstrate that the proposed joint optimization scheme outperforms metaheuristic methods, conventional beamforming approaches, and sub-schemes with disabled variables. The results also verify the effectiveness of the optimization order, where prioritizing position optimization yields the highest rate performance. The results also reveal that a compact FSIM with a small inter-layer thickness is preferable, as increasing the thickness weakens inter-layer coupling and degrades the achievable sum rate. Furthermore, the FSIM with full-boundary position reconfigurability achieves the largest spatial degrees of freedom and the highest sum-rate among existing fluid structures with restricted movement, partial FEs, element grouping, patch-based boundaries, and discrete candidate positions.

\bibliographystyle{IEEEtran}
\bibliography{IEEEabrv}
\end{document}